\documentclass[12pt,a4paper]{article}
\usepackage[textwidth=450pt, textheight=660pt]{geometry}
\usepackage{amsmath,amsthm,amssymb}
\usepackage{graphicx,tabu}
\usepackage[pagebackref=false]{hyperref}
\usepackage[active]{srcltx}
\usepackage[affil-it]{authblk}
\usepackage{wrapfig}
\usepackage[noadjust]{cite}
\usepackage{filecontents}
 \allowdisplaybreaks
 \newcommand{\be}{\begin{equation}}
\newcommand{\ee}{\end{equation}}
\newcommand{\bea}{\begin{eqnarray}}
\newcommand{\eea}{\end{eqnarray}}
\newcommand{\nn}{\nonumber}

\newcommand{\trm}{\textrm}
\newcommand{\tit}{\textit}

\newcommand{\ba}{\begin{array}}
\newcommand{\ea}{\end{array}}
\newcommand{\bfig}{\begin{figure}}
\newcommand{\efig}{\end{figure}}

\newcommand{\hs}{\hspace}
\newcommand{\bi}{\begin{itemize}}
\newcommand{\ei}{\end{itemize}}
\newcommand{\ben}{\begin{enumerate}}
 \newcommand{\een}{\end{enumerate}}

\begin{document}

  \vspace{2cm}

  \begin{center}
    \font\titlerm=cmr10 scaled\magstep4
    \font\titlei=cmmi10 scaled\magstep4
    \font\titleis=cmmi7 scaled\magstep4
  {\bf   Phononic Casimir  corrections  for Graphene resonator}

    \vspace{1.5cm}
     \noindent{{\large Y. Koohsarian ${}^a$ \footnote{yo.koohsarian@mail.um.ac.ir}, K. Javidan ${}^a$ \footnote{Javidan@um.ac.ir},
  A. Shirzad  ${}^{b,c}$ \footnote{shirzad@ipm.ir}}} \\
     ${}^{a}$ {\it Department of Physics, Ferdowsi University of Mashhad \\
       P.O.Box  91775-1436, Mashhad, Iran} \\
   ${}^b$ {\it Department of Physics, Isfahan University of Technology \\
       P.O.Box 84156-83111, Isfahan, Iran,\\
      ${}^c$  School of Physics, Institute for Research in Fundamental Sciences (IPM),\\
       P. O. Box 19395-5531, Tehran, Iran} \\

  \end{center}
  \vskip 2em

\begin{abstract}
By calculating a  Casimir energy for the acoustic phonons  of Graphene, we   find some temperature-dependent corrections for the pretension of a   Graphene sheet suspended on a trench. We obtain values of the order of few mN/m for these corrections  in fully as well as doubly clamped Graphene on a narrow trench with one nanometer width, at room temperature. These values are considerable compared to the experimental values, and can increase the fundamental resonance frequency of the Graphene. The values of these corrections increase by increasing the temperature, and so they can be utilized for tuning the Graphene pretension.
\end{abstract}

 \textbf{Keywords} \\    Graphene sheet,  Acoustic  Phonons,  pretension, Resonance frequency, Casimir energy,  Temperature correction, Nanoelectromechanical systems.

\section{Introduction} \label{sec-1}
   Nowadays Casimir forces, as important macroscopic manifestations of the quantum zero-point energies, have been investigated   for various systems, see e.g. \cite{CR1,CR2,CR3} as reviews. Primarily the Casimir forces, as attractive  interactions between some chargeless conductors being microscopically apart, are arisen from the quantum zero-point energies of electromagnetic modes   discretized in the presence of the metallic boundary walls. However   one can generally calculate a Casimir energy  by finding a finite value for the total   zero-point   energy of   any system with a  mode spectrum discretized by some appropriate physical boundaries.    As we  know,  the Casimir forces  are significant in microscopic distances, so they can imply magnificent results for tiny scale systems, see e.g. \cite{CNR1,CNR2,CNR3} as reviews.     
   
  Graphene, a one-atom-thick layer of Carbon atoms covalently bound in a honeycomb hexagonal lattice, as the thinnest known material,   has represented some remarkable mechanical and electrical properties \cite{GR1,GR2} and is expected to  provide a vast area of promising applications in  nano scale devices  known as the nanoelectromechanical systems (NEMS), see e.g. \cite{NEMS1,NEMS2,NEMS3}. As we know, a  NEMS is prototypically a nano-resonator, i.e. a nanoscale beam of an appropriate material   being actuated by an applied  external  force, to vibrate in a particular resonance mode, see e.g. \cite{NEMS4,NEMS5,B-th}. 
  
   It has been shown that the fundamental resonance frequency of a monolayer Graphene sheet suspended over a trench, is highly influenced by a significant  pre-tension induced specifically by the  strong van der Waals forces which clamp the Graphene to the sidewalls of the trench, see Refs. \cite{GV1,GV2,GV3,GV4,GT1,GT2,GT3}. In a previous work \cite{p2} we have shown that  the    zero-point energy of  the Graphene  acoustic phonons can have a considerable influence on this pre-tension.  In fact, in that work, by regarding the acoustic modes of a membrane as the oscillation modes of a massless bosonic field  having two polarizations (corresponding to transverse and longitudinal vibrations), and being confined in a rectangle cavity (corresponding to a rectangular trench), we have obtained a ``phononic'' Casimir energy for a  suspended Graphene sheet being fully clamped to the sidewalls of a rectangular trench. Then by  assigning a Casimir force to this   energy, and interpreting it as a correction to the initial tension of the membrane (see also Ref. \cite{p1}), we have obtained some temperature-dependent correctional terms for the Graphene pretension. We have shown that for a monolayer  Graphene sheet suspended on a narrow rectangular trench  with  a width $\sim \trm{nm}$ and  length $\sim \mu \trm{m}$,  at room temperature,   this correction   can be noticeable ($ \sim10^{-1} \trm{mN/m}$) compared to the experimental values of the Graphene pretension ($\sim 1 \textendash 10 \trm{mN/m}$) given in Refs. \cite{GT1,GT2,GT3}. Note that  this ``phononic''  Casimir forces  are different  from the conventional  ``photonic'' Casimir  interactions e.g.  between  two Graphene sheets or   between a Graphene sheet  and  a substrate,  which are arisen actually from the  zero-point  oscillations of the   electromagnetic field. As shown in  Refs. \cite{CEMG1, CEMG2, CEMG3,CEMG4} these electromagnetic Casimir interactions also can be significantly strong for the Graphene sheets at room temperature.
   
   In this work     we find  a similar correction for a Graphene sheet being  doubly clamped to the sidewalls of a trench. We also  explore the influence of  flexural   acoustic modes on this correction for fully as well as doubly clamped Graphene.  As we know, the flexural modes, unlike the longitudinal/transverse modes, are described by a non-linear dispersion relation, so here we may need to partly change some calculations.  In the next section, having the flexural modes been included, we find the phononic Casimir correction to the Graphene pretension, by using similar approach as in the previous work \cite{p2}. In the section 3, using a rather similar approach we calculate these corrections for the case of a  doubly clamped Graphene.    We see  that  for doubly as well as fully clamped Graphene sheet over a trench of few nanometer width, at room temperature,  these   corrections can be significant  $\sim \trm{mN/m}$ , one order of magnitude larger than that of our previous work.

\section{Fully clamped Graphene}
 
\begin{figure}[t!]
\centering
\includegraphics[width=8 cm, height=3 cm]{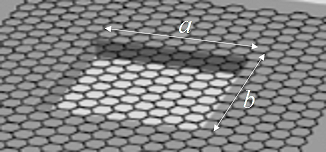}
\caption{Fully clamped Graphene sheet suspended on a rectangular trench}
\label{FCG}
\end{figure}

The  dispersion relations for   longitudinal (L), transverse (T), and flexural (F)  acoustic phonons in Graphene, near the center of  the    Brillouin zone, are well known, see e.g \cite{GP},

 \bea
&&  \omega_\trm{L,T} \approx v_{L,T} k \, ; \ \ \ \ \   v_\trm{L} \approx 21.3 \, km/s\, , \ \ \  v_\trm{T} \approx13.6 \,  km/s \, , \nn \\
 &&\omega_\trm{F} \approx  \alpha k^2 \, ;\ \ \ \   \alpha \approx 6.2 \times 10^{-7} m^2/s  \, ,
 \label{1}
 \eea
 in which $\omega$ and $k$ are   mode frequency and   mode wavenumber, respectively,  of the acoustic modes, and   $\upsilon$'s are the sound velocities in the Graphene sheet. For a fully clamped sufficiently tensioned few-layer Graphene sheet suspended on a rectangular trench, see Fig. \ref{FCG}, the resonance frequencies are given with a good accuracy by the familiar modenumbers  \cite{B-th,Z-th,GT1,GTT2}
 \be
  k_{n,m} =\pi  \sqrt{\frac{n^2}{a^2} +\frac{m^2}{b^2} }; \ \ \ a\geq b, \ \ \  n,m=1,2,... 
  \label{2}
\ee
in which, ``$a$" and ``$b$" are the side-lengths of  the  trench. Then regarding the mentioned  Graphene sheet, as a 2-dimensional gas of phonons in a rectangle cavity,  one can write the total  zero-point energy  (see e.g.  \cite{CR1}) of the Graphene acoustic phonons as
\bea
E_0(a,b,T) = \frac{k_{\trm{B}} T}{2} \sum_{n,m=1}^\infty \sum_{l=-\infty}^\infty \Bigg(\sum_{I=L,T}\ln\left[  l^2+ \lambda_I^2 \left(x^2 n^2 +m^2 \right) \right] \nn \\ +\ln\left[ l^2+ \lambda_\trm{F}^2 \left( n^2 x^2 +m^2 \right)^2 \right] \Bigg). 
\label{3}
 \eea
 in which $x\equiv b/a$, ``$T$'' is the Graphene temperature, $\hbar$ and $k_\trm{B}$ are the (reduced) Planck and the Boltzmann constants, respectively, and we have introduced some dimensionless parameters $ \lambda_I=\theta_I/T$ and $ \lambda_\trm{F}=\theta_\trm{F}/T$ with the effective temperatures 
  \bea
  \theta_{I} = \frac{ \hbar \upsilon_I}{2 b k_B} , \  \ \theta_\trm{F}  = \frac{\pi \hbar \alpha}{2 b^2 k_B}. 
  \label{4}
\eea
 Note that in Eq.\eqref{3} we have discarded an irrelevant constant term to make the logarithm arguments dimensionless. 
 Then as is conventional, one can rewrite Eq. \eqref{3} as a parametric integral,
 \bea
&&E_0(a,b,T)=  - \frac{k_B T}{2}\lim_{s\to 0} \frac{\partial}{\partial s}\Bigg( \sum_{I=\trm{L,R}}\int_0^{\infty} \frac{dt}{t}\frac{t^s}{\Gamma(s)}  \sum_{n,m =1}^{\infty}  \exp \left[-t   \lambda_{I}^2 (x^2 n^2+m^2)  \right] \nn \\
&& \hs{4cm}+ 2\sum_{I=\trm{L,R}} \int_0^{\infty} \frac{dt}{t} \frac{t^s}{\Gamma(s)}  \sum_{l,m,n=1}^{\infty}  \exp \left[-t( l^2 +\lambda_{I}^2[ x^2 n^2  + m^2 ]) \right] \nn \\
&& \hs{5cm}+\int_0^{\infty} \frac{dt}{t}\frac{t^s}{\Gamma(s)}  \sum_{n,m =1}^{\infty}  \exp \left[-t   \lambda_\trm{F}^2( x^2 n^2 + m^2)^2  \right]  \nn \\
&& \hspace{4cm}+ 2 \int_0^{\infty} \frac{dt}{t} \frac{t^s}{\Gamma(s)}  \sum_{l,m,n=1}^{\infty}  \exp \left[-t( l^2 + \lambda_\trm{F}^2[ x^2 n^2 + m^2]^2)  \right]\Bigg).
  \label{5}
\eea
The  first two lines of the above equation, i.e. the contributions of the longitudinal and the transverse acoustic modes, has been calculated exactly  in Ref. \cite{p2};  Aside from    contribution of the unbounded space which must be subtracted, the second line of the above equation contains  some   exponential and/or Bessel function terms, which for sufficiently small $\lambda$'s can be neglected with a good degree of accuracy. By a similar discussion, one can neglect the  fourth line of the above equation. Therefor for sufficiently small $\lambda$'s, that is, for sufficiently large temperatures compared to the effective temperatures  \eqref{4}, the zero-point energy \eqref{5} can be approximated by its first and   third lines;
 \bea
&&E_0(a,b,T) \approx  - \frac{k_B T}{2}\lim_{s\to 0} \frac{\partial}{\partial s}\Bigg( \sum_{I=\trm{L,R}}\int_0^{\infty} \frac{dt}{t}\frac{t^s}{\Gamma(s)}  \sum_{n,m =1}^{\infty}  \exp \left[-t   \lambda_{I}^2 (x^2 n^2+m^2)  \right]  \nn \\
&& \hs{2cm}+\int_0^{\infty} \frac{dt}{t}\frac{t^s}{\Gamma(s)}  \sum_{n,m =1}^{\infty}  \exp \left[-t   \lambda_\trm{F}^2( x^2 n^2 + m^2)^2  \right]  \Bigg); \ \ \  T\gg \theta_\trm{L,T,F}
  \label{6}
\eea
Note that at room temperature ($\approx 300 K$), the above approximation is acceptable for $b\gtrapprox 0.5 \trm{nm}$, see Eq. \eqref{4}   (having $\hbar \approx 1.05 \times 10^{-34} m^2 kg/s$ and $k_B \approx 1.38 \times 10^{-23} m^2 kg/s^2 K$). Now  the first line of the above equation can be simplified by using the relation;
\be
 \sum_{ n=1}^{\infty} \exp \left[-\alpha n^2  \right]=-\frac{1}{2}+ \frac{1}{2 } \sqrt{\frac{\pi}{\alpha}}+  \sqrt{\frac{\pi}{\alpha}} \sum_{n=1}^{\infty} \exp{\left[-\frac{\pi^2 n^2}{ t\alpha} \right]}; \ \ \ \alpha>0,
  \label{7}
\ee
which can be obtained directly by using the Poisson summation formula (see e.g. \cite{math1}),
\be
\sum_{n=1}^{\infty} f(n)= -\frac{f(0)}{2} + \int_0^{\infty} f(x) dx +2 \sum_{n=1}^{\infty} \int_0^{\infty} f(x) \cos(2\pi nx)dx. \label{8}
\ee
 Then  by utilizing the Riemann zeta function $ \zeta (s)=\sum_{n=1}^{\infty} n^{-s}$, and the integral relation
\be
\int_0^{\infty}  t^r \exp\left[-x^2 t-y^2/t \right]dt=2(x/y)^{-r-1}K_{-r-1}(2xy),
 \label{9}
\ee
with the Modified Bessel function
\be
K_\nu(z) = \frac{(z/2)^\nu \Gamma(1/2)}{\Gamma(\nu+1/2)} \int_1^\infty e^{-zt} \left(t^2-1 \right)^{(2\nu-1)/2} dt,
\label{10}
\ee
one can find
\bea
&&\int_0^{\infty} \frac{dt}{t}\frac{t^s}{\Gamma(s)}  \sum_{n,m =1}^{\infty}  \exp \left[-t   \lambda_{I}^2 (x^2 n^2+m^2)  \right]=\nn \\ 
&& \hspace{3cm} -\frac{\zeta (2 s)}{2(x \lambda _{I})^{2 s} } +\frac{\sqrt{\pi } }{ 2 \lambda _{I}} \frac{ \Gamma \left(s-1/2\right)}{ \Gamma (s)}\frac{ \zeta (2 s-1)}{(x \lambda _{I})^{ 2 s-1}} \nn \\ 
&& \hspace{3 cm} +2 \frac{\sqrt{\pi}}{\lambda _{I} \Gamma(s)}  \sum_{n,m=1}^{\infty} \left( \frac{n x \lambda_{I}}{m \pi/\lambda_{I}}\right)^{ -s+1/2} K_{ -s+1/2}\left( 2 \pi  n m x \right)
\label{11}
\eea
But the second line of Eq. \eqref{6}, which comes from the contribution of the flexural modes, can not be simplified, in its present form, by directly using a   relation of the form \eqref{7}. However we can   re-parametrize  the mentioned expression as
 \bea
\int_0^{\infty} \frac{dt}{t}\frac{t^s}{\Gamma(s)} \exp \left[-t   \lambda_\trm{F}^2( x^2 n^2 + m^2)^2  \right] &=&   \left[  \lambda_\trm{F}^2( x^2 n^2 + m^2)^2  \right]^{-s} \nn \\
 &=& \lambda_\trm{F}^{-2s}  \int_0^{\infty} \frac{dt}{t}\frac{t^{2s}}{\Gamma(2s)} \exp \left[-t    ( x^2 n^2 + m^2)  \right]. 
 \label{12}
 \eea
Now one can apply the summation \eqref{7} to the $m$-sum in the second line of  the above equation, and after some calculations similarly as carried out for the first line of Eq. \eqref{6}, we obtain
\bea
&&\int_0^{\infty} \frac{dt}{t}\frac{t^s}{\Gamma(s)}  \sum_{n,m =1}^{\infty}  \exp \left[-t   \lambda_\trm{F}^2( x^2 n^2 + m^2)^2  \right] = \nn \\ 
&& \hspace{3 cm} \lambda_\trm{F}^{-2s} \Bigg[-\frac{\zeta (4 s)}{2 x ^{4 s} }+\frac{\sqrt{\pi } }{ 2  } \frac{ \Gamma \left(2s-1/2\right)}{ \Gamma (2s)}\frac{ \zeta (4s-1)}{x  ^{ 4 s-1}} \nn \\ 
&& \hspace{3 cm} +2 \frac{\sqrt{\pi}}{  \Gamma(2s)}  \sum_{n,m=1}^{\infty} \left( \frac{n x  }{m \pi }\right)^{ -2s+1/2} K_{ -2s+1/2}\left( 2 \pi  n m x \right) \Bigg]
\label{13}
\eea
Eventually after a rather long calculation by using
 \bea
&& \lim_{s \to 0} \frac{\partial}{\partial s} \frac{g(s)}{\Gamma(s)}=g(0), \nn \\
 &&  \lim_{s\to 0} \, \frac{\frac{\partial}{\partial s} \trm{K}_{ -s-r} (x)}{\Gamma (s)}=0 ; \ \ \ r\geq 0
\label{14}
\eea
in Eq. \eqref{6} we obtain an expression for the phononic Casimir energy of  the Graphene sheet;
\bea
&&E_\trm{C}(a,b,T)  \approx -k_B T \Bigg( \frac{ \pi}{ 6} x  - \ln x-\frac{\ln(\lambda _\trm{L}\lambda _\trm{T}\lambda _\trm{F} )}{4}   \nn \\
&& \hspace{4cm}+2\sum_{n,m=1}^{\infty} \frac{\exp{\left(- 2  m n\pi x \right)}}{m}  \Bigg) \, ;  \ \ \ x\equiv b/a, \ \ \ \lambda_{\trm{L},\trm{T},\trm{F}} \ll 1
 \label{15}
\eea
Then the phononic Casimir forces can be introduced as
\bea
&&F_{\textrm{C},a} (a,b,T)  \equiv- \frac{\partial  E_C }{\partial a} \approx   \frac{ k_B T}{a} \left(1-\frac{\pi}{6} \frac{b}{a} +\frac{b}{a}  S\left(\frac{b}{a}\right)   \right) \nn \\ 
&&F_{\textrm{C},b} (a,b,T)  \equiv - \frac{\partial  E_C }{\partial b} \approx \frac{ k_B T}{a} \left( \frac{\pi}{6}-  S\left(\frac{b}{a}\right)  \right)
 \label{16} 
\eea
in which
\bea
S(x) \equiv  \sum_{n,m=1}^{\infty} 4 n\pi\exp{\left(- 2  m n\pi x \right)}; \ \  x \leq 1.
\label{17}
\eea
One can see that the   forces \eqref{16} are just twice the Corresponding forces in Ref. \cite{p2}, where the contribution of the flexural modes had not been taken into account. Hence the contribution of the flexural modes, to the phononic Casimir forces of the Graphene sheet, equals just twice the   contribution of transverse/longitudinal modes.

Now, as   shown in Ref. \cite{p2}, the above phononic Casimir forces can be interpreted as corrections to the pretension of Graphene sheet;
\bea
\tau_a=\tau_0 - \frac{1}{b} F_{\textrm{C,a}} \nn \\
\tau_b=\tau_0 - \frac{1}{a} F_{\textrm{C,b}} 
\label{18}
\eea
in which, $\tau_0$ is the Graphene pretension. As a result one can write
\bea
\tau_{a,b} (a,b,T) \approx \tau_0 + \Delta_{a,b} (a,b,T)   
\label{19}
\eea
 with
 \bea
&& \Delta_a (a,b,T)\equiv - \frac{ k_B T}{b^2}  \left[   \frac{1}{2}\frac{b}{a}-\frac{\pi}{ 12} \left(\frac{b}{a} \right)^2+ \left(\frac{b}{a} \right)^2 S \left(\frac{b}{a}\right) \right]    \nn \\
&&  \Delta_b(a,b,T) \equiv  \frac{ k_B T}{b^2} \left[- \frac{\pi}{ 12} \left(\frac{b}{a} \right)^2+ \left(\frac{b}{a} \right)^2 S \left(\frac{b}{a}\right) \right] 
  \label{20}
\eea
  By numerical computations, one can see that $S(x) \geq S(1) \approx 0.02$, so (the absolute value of ) $\Delta_a$  as well as $\Delta_b$ increases by decreasing the ratio $b/a$. However,  $\Delta_a$ is always  negative  i.e. subtractive to  the pretension, while  $\Delta_b$ is negative for $b/a\gtrapprox 0.5$, and positive, i.e. additive to the pretension, for $b/a\lessapprox 0.5$. 

For a narrow trench ($b/a\ll1$) we can write
\bea
\tau_{a,b} (a,b,T) \approx \tau_0 \mp   \frac{ k_B T}{a^2} S \left(\frac{b}{a}\right); \ \ \ b/a\ll1
  \label{21}
\eea
in which ``$+$'' (``$-$'') corresponds to the index ``$a$'' (``$b$''). For instance, for a suspended fully clamped Graphene sheet on a  narrow  trench with $a=1 \mu\trm{m}$, $b=1\trm{nm}$, at room temperature ($\approx 300 \trm{K}$), using $S(0.001) \approx 5.2 \times 10^5$ one obtains 
\be
\tau_{a,b} (1\mu \trm{m}, 1 \trm{nm}, 300 \trm{K}) - \tau_0 \approx \mp 2.1 \ \trm{mN/m} .
\label{22}
\ee
This is a considerable value compared to the experimental values for the Graphene pretension  $ \sim 1 \textendash 10 \ \trm{mN/m}$, see Refs. \cite{GT1,GT2,GT3}. Note that the  above correction is one order of magnitude larger  then that of Ref. \cite{p2}, which is a consequence of  taking the contribution of flexural modes into account, as we mentioned before.   However for rectangular trenches  having widths ($b$)  larger than few nanometer,  one can see that the   correction \eqref{21} would be negligible.

The change in the Graphene tensions ($\Delta \tau_{a,b}$) due to the change $\Delta T$ in the Graphene temperature, can be written as
\bea
\Delta \tau_{a,b} \approx \mp   \frac{ k_B \Delta T}{a^2} S \left(\frac{b}{a}\right) 
\label{23}
\eea
Note that these corrections generally break the (tensional) isotropy of the Graphene ($\tau_a \neq \tau_b$). As a result, the  fundamental resonance frequency ($f_{11}$) of the Graphene sheet \cite{Z-th,GTT2}, changes by temperature as
\bea
 \Delta f_{11}^2 & =&  \frac{\sigma}{4\alpha \rho_0}\left( \frac{\Delta\tau_a}{a^2} +\frac{\Delta \tau_b}{b^2}\right) \nn \\
&\approx&  \frac{ \sigma k_B}{4 \alpha \rho_0 } \frac{ S \left(b/a\right)}{a^2 b^2} \Delta T.
  \label{24}
\eea
in which, $ \Delta f_{11}^2 $ is the change in the square-value of  $f_{11}$,   $\sigma = 340 \trm{N/m}$ and $\rho_0 = 7.4\times 10^{-7} \textrm{kg/m}^2$  are the in-plane stiffness and the  density of the mono-layer Graphene, respectively, and $\alpha=\rho_\trm{total}/\rho_0 \sim 1\textendash10$  is the adsorbed mass coefficient, see  also  \cite{B-th,GT1,GT2}).

 \section{Doubly clamped Graphene}
\begin{figure}[t!]
\centering
\includegraphics[width=9 cm, height=3cm]{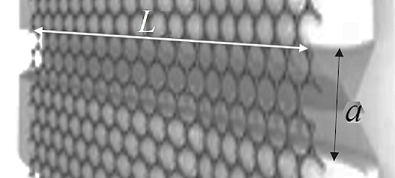}
\caption{Doubly clamped Graphene suspended on a narrow trench ($a\ll L$)}
\label{DCG}
\end{figure}

For a doubly clamped Graphene sheet suspended on a narrow trench as in Fig. \ref{DCG}, the phononic zero-point energy \eqref{3} takes the form
\bea
E_0(a,b,T) = \frac{k_{\trm{B}} T}{2}\int_0^\infty dk \sum_{n =1}^\infty \sum_{l=-\infty}^\infty \Bigg(\sum_{I=L,T}\ln\left[  l^2+ \lambda_I^2 \left(y^2 k^2 +n^2 \right) \right] \nn \\ +\ln\left[ l^2+ \lambda_\trm{F}^2 \left( y^2 k^2 +n^2 \right)^2 \right] \Bigg). 
\label{25}
 \eea
 in which $y\equiv a/L$,  $ \lambda_I=\theta_I/T$ and $ \lambda_\trm{F}=\theta_\trm{F}/T$ with the effective temperatures 
  \bea
  \theta_{I} = \frac{ \hbar \upsilon_I}{2 a k_B} , \  \ \theta_\trm{F}  = \frac{\pi \hbar \alpha}{2 a^2 k_B}. 
  \label{26}
\eea
As a result similar to Eq. \eqref{6} one can  write
 \bea
&&E_0(a,b,T) \approx  - \frac{k_B T}{2}\lim_{s\to 0} \frac{\partial}{\partial s}\Bigg( \sum_{I=\trm{L,R}}\int_0^{\infty} \frac{dt}{t}\frac{t^s}{\Gamma(s)}  \sum_{n=1}^{\infty}  \exp \left[-t   \lambda_{I}^2 (y^2 k^2+n^2)  \right]  \nn \\
&& \hs{2cm}+\int_0^{\infty} \frac{dt}{t}\frac{t^s}{\Gamma(s)}  \sum_{n =1}^{\infty}  \exp \left[-t   \lambda_\trm{F}^2( y^2 k^2 + n^2)^2  \right]  \Bigg); \ \ \  T\gg \theta_\trm{L,T,F}
  \label{27}
\eea
with  $k$'s  as  the  wavenumbers  of  continuous acoustic modes propagating parallel to the trench.   Note that at room temperature, the above approximation is valid for $a\gtrapprox 0.5 \trm{nm}$. The contribution of  T- and L-modes (in the first line of the above equation) can be directly simplified by using a Gaussian integral for ``$k$'' and a zeta regularization for the $n$-sum;
\bea
 \int_0^{\infty} \frac{dt}{t}\frac{t^s}{\Gamma(s)}  \sum_{n=1}^{\infty}  \exp \left[-t   \lambda_{I}^2 (y^2 k^2+n^2)  \right] 
  &=& \frac{\sqrt{\pi}}{ 2y \lambda_I^{2s}} \int_0^{\infty} \frac{dt}{t}\frac{t^{s-1/2}}{\Gamma(s)}  \sum_{n=1}^{\infty}  \exp \left[-t    n^2  \right] \nn \\
&=&  \frac{\sqrt{\pi}}{2y \lambda_I^{2s}} \frac{\Gamma(s-\frac12)}{\Gamma(s)} \zeta(2s-1)
\label{28}
\eea
Similar calculation can be done for the contribution of  F-modes (in the second line of Eq. \eqref{27}), after a reparametrization;
\bea
\int_0^{\infty} \frac{dt}{t}\frac{t^{ s}}{\Gamma( s)}  \sum_{n =1}^{\infty}  \exp \left[-t   \lambda_\trm{F}^2( y^2 k^2 + n^2)^2  \right] &=& \lambda_\trm{F}^{-2s}  \int_0^{\infty} \frac{dt}{t}\frac{t^{2s}}{\Gamma(2s)}  \sum_{n =1}^{\infty}  \exp \left[-t   ( y^2 k^2 + n^2)   \right] \nn \\
&=&   \frac{\sqrt{\pi}}{ 2y \lambda_F^{2s}} \frac{\Gamma(2s-\frac12)}{\Gamma(2s)} \zeta(4s-1)
\label{29}
\eea
Finally substituting Eq. \eqref{28} and \eqref{29} into the zero-point energy \eqref{27} and using the relations \eqref{14}, one can obtain the phononic Casimir energy of the doubly clamped Graphene;
\be
E_\trm{C} (a,T) \approx -\frac{\pi k_\trm{B} T}{6} \frac{L}{a} ; \ \ \   T\gg \ \theta_\trm{L,T,F}, \ \ a\ll L
\label{30}
\ee
As a results, similarly as  the previous section, the corrected pretension of the Graphene can be given as
\bea
\tau (a,T)&= & \tau_0 -\frac1L F_\trm{C} \nn \\
&\approx& \tau_0+\frac{\pi k_\trm{B} T}{6 a^2}.
\label{31}
\eea
As is seen, the correction term is always additive. For a narrow trench having the width $a=1\trm{nm}$, at room temperature, one obtain
\be
\tau (1\trm{nm},300 \trm{K})- \tau_0 \approx 2.2 \ \trm{mN/m}
\label{32}
\ee
which is again considerable compared to the experimental values ($\tau_0 \sim 1\textendash 10 \trm{mN/m}$) \cite{GT1,GT2,GT3}. Note that the above correction for the doubly clamped Graphene is approximately equal to the corresponding correction for the Fully clamped Graphene, see Eq. \eqref{22}.  The tension change in terms of temperature change can be written as
\bea
\Delta \tau \approx  \frac{\pi k_\trm{B} \Delta T}{6 a^2}.
\label{33}
\eea
 which  changes the first resonance frequency of the doubly clamped Graphene \cite{Z-th,GTT2} as
 \bea
 \Delta f_{11}^2 & =&  \frac{\sigma}{4\alpha \rho_0} \Delta\tau \nn \\
&\approx&  \frac{ \pi \sigma }{24 \alpha \rho_0  } \frac{k_B \Delta T}{a^2}.
  \label{34}
\eea

\section{Concluding remarks}
We have obtained the  Casimir energy for acoustic phonons in fully as well as doubly clamped Graphene sheet suspended on a trench, at finite temperature.  We have introduced Phononic Casimir forces as changes in the mentioned Casimir energy due to infinitesimally changing the trench sidelengths, and  interpreted these Caimir forces as corrections to the Graphene pretension. Then by numerical computations, for narrow trenches of $1$nm width at room temperature, we have obtained values  in the order of few mN/m for these corrections,  which are considerable compared to the experimental values  of  Refs. \cite{GT1,GT2,GT3}. It is also interesting to note that the values of  these corrections increase by increasing the temperature, see Eqs. \eqref{21} and \eqref{33}, while the Graphene pretension decreases by increasing the temperature \cite{GTT1,GTT2,GTT3}. Hence   these corrections can be even more  considerable by increasing the temperature, and so they can be utilized for tuning the Graphene pretension.

\subsection*{ Acknowledgment}
We thank  S. A.  Jafari for his valuable comments, and S.  Qolibikloo for his helps during the numerical computations.

%%%%%%%%%%%%%%%%%%%%
\end{document}